\documentclass[11pt]{amsart}

\usepackage{amssymb}
\usepackage{amsmath}
\usepackage{enumerate}
\usepackage{mathtools}
\usepackage{color}
\usepackage[T1]{fontenc}
\begin{document}
\title[Sparse PIC]{Sparse grid techniques for particle-in-cell schemes}

\author{L F Ricketson, A J Cerfon}
\address{$^1$Courant Institute, New York University, 251 Mercer St, New York, NY 10012}

\begin{abstract}
We propose the use of sparse grids to accelerate particle-in-cell (PIC) schemes.  By using the so-called `combination technique' from the sparse grids literature, we are able to dramatically increase the size of the spatial cells in multi-dimensional PIC schemes while paying only a slight penalty in grid-based error.  The resulting increase in cell size allows us to reduce the statistical noise in the simulation without increasing total particle number.  We present initial proof-of-principle results from test cases in two and three dimensions that demonstrate the new scheme's efficiency, both in terms of computation time and memory usage.
\end{abstract}

%\keywords{sparse grid, particle-in-cell}
%\submitto{\PPCF}

\maketitle

\section{Introduction}
The particle-in-cell (PIC) method has been a standard tool in the simulation of kinetic plasmas for over 50 years \cite{birdsall1969clouds,buneman1963computer,dawson1962one,langdon1970theory,morse1970multidimensional}, and continues to be popular today - see e.g.\ \cite{birdsall2004plasma,chen2013energy,fubiani2015developpment, garrigues2015characterization,philippov2014ab,taitano2013development,tenbarge2014collisionless}, among many other references.  Its numerous desirable features explain in large part its success - it is conceptually simple, readily parallelizable, relatively robust, and can incorporate a wide variety of physical phenomena.  Even so, PIC simulations of the complex, three-dimensional systems that arise in modern plasma physics applications still require many hours on a massively parallel machine \cite{exum2013ppps, fiuza2013ion, fonseca2013exploiting, fubiani2015developpment, wang2015modern}.  

There are several reasons for these enormous run times, but prominent among them is the statistical error introduced by the particle representation.  Good statistical resolution requires many particles \textit{per cell}, and in two or three dimensions, the number of cells may be very large in order to properly resolve the features of the system under study.  The combination of these two requirements forces the total number of particles to be overwhelmingly large, making for a computation that requires both huge CPU and memory resources.  

The result of this interaction between the grid and particles requirements is an algorithm with two of the worst features of both particle and continuum methods:  One is not only saddled with the unavoidably slow convergence of a particle scheme, but also complexity that grows exponentially with dimension, like a continuum method.  While the reduction in dimension - from six to three - makes this a worthwhile sacrifice in many contexts, further acceleration of PIC simulations is clearly of great practical interest.  

In this paper, we propose the use of so-called `sparse grids' in concert with PIC to obtain a scheme whose complexity is nearly independent of dimension.  The idea of a sparse grid is not new, having been studied extensively in the applied mathematics community \cite{bungartz2004sparse,garcke2001data,griebel1998adaptive,griebel1990combination}.  It has also seen some use in continuum codes for plasma applications \cite{ali2015fault,guo2016sparse,heene2013load,hinojosa2015towards,kowitz2013sparse,kowitz2012combination,pfluger2014exahd}.  

While most of the sparse grid literature deals with a hierarchical-basis representation \cite{bungartz2004sparse}, here we will make use of an alternative characterization called the `combination technique' \cite{griebel1990combination}.  
Specifically, a function - e.g. charge density or current - is approximated on a variety of different grids, each of which has a different resolution in each coordinate direction.  By combining these approximations intelligently, one can achieve accuracy close to that of a well-resolved regular grid, but at dramatically reduced cost.  

Crucially for PIC, the combination technique grids have very large cells relative to a comparable regular grid.  This improves statistical resolution by increasing the number of particles \textit{per cell} without increasing the overall particle number.  The use of sparse grids can thus accelerate not only the computation of the electromagnetic fields, but also the particle operations, which typically dominate the computation and storage requirements.  The key to achieve this end is to reinterpret the particle shape function typically used in PIC as an approximation to a delta function rather than a physical charge density.  

The remainder of the article is structured as follows.  In section 2, we present the necessary background on the combination technique and particle-in-cell methods in general.  In section 3, we present our reinterpretation of PIC methodology and the use of sparse grids that it enables.  In section 4, we proceed to results of numerical experiments in 2- and 3-D.  In section 5, we discuss our results, prospects for future work, and conclusions.  

%%%%%%%%%%%%%%%%%%%%%%%%%%%%%%%%%%%%%%%%%%%%%%%%%%%%%%%%%%

\section{Prerequisites}
In this section, we establish the necessary background on the combination technique for sparse grids and PIC schemes.  

\subsection{Sparse grid combination technique}
While the combination technique can be applied to problems of any dimensionality greater than one, we will discuss the key ideas in the context of two dimensional simulations for the simplicity of the presentation.  We refer the interested reader to previous work \cite{griebel1990combination,griebel1992combination,hegland2007combination,reisinger2012analysis} for a more complete treatment.  Furthermore, even though the sparse grids are not limited to approximation schemes based on multi-linear bases \cite{achatz2003higher,bungartz2004higher,bungartz2004sparse}, we restrict for simplicity our presentation here to multilinear bases, and leave the possibility of PIC schemes with higher-order bases on sparse grids to future work. 

Consider, then, a function $u(x,y)$ on the unit square $[0,1] \times [0, 1]$ that we wish to approximate on a rectangular grid with cell width $h_x$ and height $h_y$. Two simple circumstances of this type are (a) $u$ is the solution of an elliptic PDE that we solve with bilinear finite elements, and (b) $u$ is known at the grid points and we wish to approximate it on the entire domain via bilinear interpolation.  For these and many other approximation schemes based on a piecewise bilinear representation of a smooth function, the error between the exact function $u$ and its approximation $\mathfrak{u}$ at a particular point can be written has
\begin{equation} \label{uerr}
	u(x,y) - \mathfrak{u}(x,y) = C_1(h_x) h_x^2 + C_2(h_y) h_y^2 + C_3(h_x,h_y) h_x^2 h_y^2.
\end{equation}
The $C_i$ above are functions with a uniform upper bound.  Importantly, by allowing the $C_i$ to be arbitrary functions of their arguments, they can be chosen so that (\ref{uerr}) is not simply the leading terms in an asymptotic expansion, but rather the \textit{exact} error \cite{griebel1990combination}.  

In the absence of additional information about $u$, one typically chooses $h_x = h_y = h$ and finds an error of $O(h^2)$.  In the best case scenario, the computational complexity of the scheme scales linearly in the number of grid points, giving a complexity that scales like $O(h^{-2})$.  Thus, the complexity $\kappa$ is related to the error $\varepsilon$ by 
\begin{equation}
	\kappa \sim \varepsilon^{-1}.
\end{equation}
In arbitrary dimension $d$, the best case computational complexity scales as $O(h^{-d})$, meaning the above generalizes to $\kappa \sim \varepsilon^{-d/2}$.

\begin{figure}[h]
\begin{center}
	\includegraphics[width=0.4\textwidth]{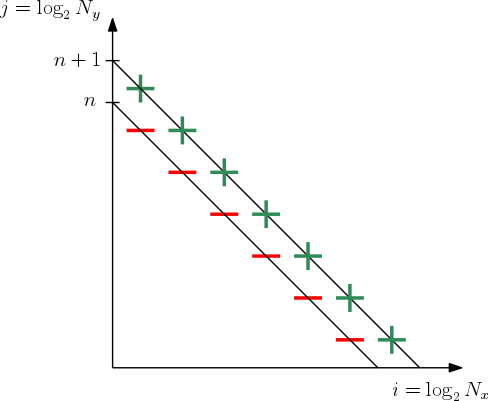}
	\caption{A graphical depiction of the combination of grids used in (\ref{combo}).  The green `+' signs represent grids that give a positive contribution, while red `$-$' signs are subtracted.  Cancellation arises from pairing neighboring grids along vertical and horizontal axes.}
	\label{combdiag}
\end{center}
\end{figure}

The combination technique improves on this circumstance by using cancellation across different grids.  Suppose the desired resolution is $h_n = 2^{-n}$ for some positive integer $n$.  Let $h_x^i = 2^{-i}$, $h_y^j = 2^{-j}$ and $\mathfrak{u}_{i,j}$ be the approximation of $u$ on the corresponding grid.  Then, consider the quantity $\mathfrak{u}_n$ defined by 
\begin{equation} \label{combo}
	\mathfrak{u}_n = \sum_{i+j=n+1} \mathfrak{u}_{i,j} - \sum_{i+j=n} \mathfrak{u}_{i,j}.
\end{equation}
In each of the sums, $i$ and $j$ are strictly positive integers.  This combination is depicted graphically in fig.\ \ref{combdiag}.

By considering figure 1 and the error formula (\ref{uerr}), we see that a great deal of cancellation occurs in computing the error corresponding to $\mathfrak{u}_n$.  In particular, for any particular $i$ between $1$ and $n-1$, a grid with horizontal spacing $h_x^i$ appears exactly once in each of the two sums in (\ref{combo}).  For those two grids, the term $C_1(h_x) h_x^2$ that appears in (\ref{uerr}) exactly cancels, because it is independent of $h_y$.  The only contribution from the $O(h_x^2)$ term thus comes from the grid with $h_x = 2^{-n}$.  An analogous argument holds in the $y$-direction.  We thus find
\begin{equation}
	u - \mathfrak{u}_n = C_1(h_n) h_n^2 + C_2 (h_n) h_n^2 + h_n^2 \left\{\frac{1}{4} \sum_{i+j=n+1} C_3(h_x^i,h_y^j) - \sum_{i+j=n} C_3(h_x^i,h_y^j) \right\},
\end{equation}
where we have used the fact that $h_x^i h_y^j = h_n/2$ when $i+j = n+1$ and $h_x^i h_y^j = h_n$ when $i+j = n$.  The expression in braces contains $2n-1$ terms which are all uniformly bounded by constants, so we find that
\begin{equation}
	|u - \mathfrak{u}_n| = O(n h_n^2) = O(h_n^2 |\log h_n|).
\end{equation}
That is, $\mathfrak{u}_n$ approximates $u$ \textit{nearly as well} as an approximate solution using $h_x = h_y = 2^{-n}$.  Even better, each grid used in the combination technique has $O(h_n^{-1})$ grid points, and there are $O(n)$ grids.  Thus, for schemes that scale with the number of grid nodes, one has $\kappa = O(h_n^{-1} | \log h_n |)$.  Using the fact that the logarithm grows slower than any polynomial, we thus find a new relationship between complexity $\kappa$ and error $\varepsilon$:
\begin{equation} \label{sparsecomplexity}
	\kappa \sim \varepsilon^{-1/2} |\log \varepsilon|^{2},
\end{equation}
At least asymptotically, one can thus achieve the same accuracy considerably faster with the combination technique than with a single regular grid.  

This idea extends naturally to higher dimensions.  In three dimensions, for example, if we let $h_z^k = 2^{-k}$, then 
\begin{equation}
	\mathfrak{u}_n = \sum_{i+j+k = n+2} \mathfrak{u}_{i,j,k} - 2 \sum_{i+j+k=n+1} \mathfrak{u}_{i,j,k} + \sum_{i+j+k = n} \mathfrak{u}_{i,j,k}
\end{equation}
gives a sparse approximation of $u$ - see \cite{griebel1990combination} for more detail.  In general, a $d$-dimensional function is represented with error $O(h_n^2 |\log h_n|^{d-1})$ and complexity $O(h_n^{-1} |\log h_n|^{d-1})$.  The scaling of complexity with error shown in (\ref{sparsecomplexity}) thus generalizes to
\begin{equation}
	\kappa \sim \varepsilon^{-1/2} |\log \varepsilon|^{2(d-1)}.
\end{equation}
For very high dimensional functions, this is an enormous savings over the $\kappa \sim \varepsilon^{-d/2}$ scaling for regular grids.  

Of course, this technique is not without its limitations.  Firstly, it requires a structured grid.  An unstructured triangular mesh, for instance, has no natural sense of $h_x$ and $h_y$ that can be refined independently, making it unclear how these ideas might be applied.  That being said, it is not necessary that the domain be rectangular.  It is only necessary that the domain can be mapped to a rectangle (or a sequence of connected rectangles \cite{bungartz2004sparse}) via some coordinate transformation.  

A second drawback is the increased regularity demanded of the function $u$.  In particular, for a regular grid, the leading order constants in the error - $C_1$ and $C_2$ - are proportional to the second derivatives $u_{xx}$ and $u_{yy}$.  In contrast, the leading term in the sparse grid error - related to $C_3$ - is proportional to the fourth-order mixed derivative $u_{xxyy}$.  As a result, the constant in front of the error scaling will often be larger for a sparse grid than a regular grid, and the predicted scaling may not even apply if $u$ is not sufficiently smooth.  Local adaptive refinement techniques have successfully mitigated this difficulty in some scenarios \cite{griebel1998adaptive}. 

%%%%%%%%%%%%%%%%%%%%%%%%%%%%%%%%%%%%%%%%%%%%%%%%%%%%%%%%%%%%%%%%%%%%%%%%%%
\subsection{Particle-in-cell}
For simplicity, we restrict our presentation to a relatively basic physical situation in which PIC applies, namely the electrostatic Vlasov equation with fixed magnetic field for the electrons, and ions assumed to form a uniform, immobile, neutralizing background:
\begin{equation} \label{ESVlasov}
\begin{split}
	&\partial_t f + \mathbf{v} \cdot \nabla_x f + q (\mathbf{E} + \mathbf{v} \times \mathbf{B} )\cdot \nabla_v f = 0, \\
	&\mathbf{E} = -\nabla \varphi, \qquad -\nabla^2 \varphi = q \int f \, d\mathbf{v} + \rho_i.
\end{split}
\end{equation}
In the above, $f(\mathbf{x},\mathbf{v},t)$ is the electron phase-space distribution, $q$ the electron charge, and $\rho_i$ the constant charge density of the ions.  

Traditionally, a PIC scheme for (\ref{ESVlasov}) is comprised of four features:
\begin{enumerate}[i.]
	\item Represent $f$ as a collection of simulated ``particles" $(\mathbf{x}_p, \mathbf{v}_p)$, which are evolved via some finite difference scheme for 
	\begin{equation} \label{ParPush}
		\dot{\mathbf{x}}_p = \mathbf{v}_p, \qquad \dot{\mathbf{v}}_p = q (\left. \mathbf{E} + \mathbf{v} \times \mathbf{B} )\right|_{\mathbf{x} = \mathbf{x}_p}
	\end{equation}
	
	\item Assign to each particle a charge density $S(\mathbf{x} - \mathbf{x}_p)$, and approximate the overall electron charge density at grid points $\mathbf{x}_k$ via
	\begin{equation} \label{PICparadigm}
		\rho_e(\mathbf{x}_k) \approx \varrho_e (\mathbf{x}_k) \coloneqq \sum_p S(\mathbf{x}_k - \mathbf{x}_p).
	\end{equation}
	
	\item Use some grid-based Poisson solver to compute $\varphi$ by solving $-\nabla^2 \varphi = \rho_e + \rho_i$, then evaluate $\mathbf{E}$ by numerical differentiation.  
	
	\item Evaluate $\mathbf{E}$ at the particle positions $\mathbf{x}_p$ by interpolation.  Use this to repeat step i.
\end{enumerate}
One loop through each of these four phases represents a single time step.  We will denote the total number of simulated particles by $N_p$.  For simplicity, we will assume the spatial domain is rectangular and the use of a regular Cartesian grid with cell width $h_x$, $h_y$, and $h_z$ in the coordinate directions.

Let us consider the simplest practical PIC scheme, which uses a leapfrog or Boris scheme for the particle push and a second-order finite element scheme for the field solve.  Typically, the grid spacing is approximately equal in each coordinate direction - call it $h$.  The overall error $\varepsilon$ of the scheme has the following scalings \cite{birdsall2004plasma}:
\begin{equation} \label{stderrscalings}
	\underbrace{\varepsilon \sim \Delta t^2}_{\substack{\text{Time-stepping}\\\text{error}}}, \qquad \underbrace{\varepsilon \sim h^2}_{\substack{\text{Grid}\\\text{error}}}, \qquad \underbrace{\varepsilon \sim (N_p h^{d_x})^{-1/2}}_{\substack{\text{Particle sampling}\\\text{error}}}.
\end{equation}
Observe that an ideal particle method has a sampling error that scales like $N_p^{-1/2}$, but PIC's sampling error scales not with the total particle number, but the number of particles \textit{per cell}.  A small grid error requires $h^2 \ll 1$, but small sampling error requires not only $N_p \gg 1$ but also the more onerous condition $N_p \gg h^{-d_x}$.  

It is also evident that $N_p$ must grow rapidly as $d_x$ increases.  This point is further emphasized by looking at the complexity of the particle based operations - the particle push and interpolation to/from the grid.  The complexity $\kappa$ scales with $N_p / \Delta t$.  A simple calculation leads from (\ref{stderrscalings}) to
\begin{equation}
	\kappa \sim \varepsilon^{-(2.5 + d_x/2)}.
\end{equation}
Evidently, a 2-D PIC scheme is more expensive than a 1-D scheme by a factor of $1/\sqrt{\varepsilon}$, while the complexity of a 3-D scheme is increased by $1/\varepsilon$ compared to 1-D.  

\subsubsection{The Shape Function}
This work is concerned with the use of sparse grid ideas, which necessitate a reinterpretation of the shape function $S$ - which maps the particle properties onto the grid - and the errors associated with that process.  To help clarify our novel conception of $S$, we first describe here the usual interpretation.

While a variety of choices exist, we will assume for simplicity that $S$ is built from the so-called \textit{tent function} $\tau$, defined by
\begin{equation}
	\tau(x) = \left\{ \begin{array}{ccr} 1 - |x| & : & |x| \leq 1 \\ 0 & : & |x| > 1 \end{array} \right.
\end{equation}
$S$ is then given by
\begin{equation}
	S(\mathbf{x}) = \frac{Q}{N_p} \prod_{i=1}^{d_x} \frac{\tau \left(x_i / h_i \right)}{h_i},
\end{equation}
where $h_i$ is the cell width in the $i^\textrm{th}$ coordinate direction.  The scaling factor $Q$ is interpreted as the total (electron) charge in the system, and this choice of $S$ fixes that value independent of $N_p$.  Physically, if the total number of \textit{physical} particles in the system is $\mathcal{N} = Q / q$, each simulated particle is interpreted as a \textit{macroparticle} with charge $q \mathcal{N} / N_p$ that represents \textit{many} physical particles.  

With these standard concepts set, we are now ready to revisit the PIC method in the context of sparse grids and the combination technique.

\section{Merging PIC with Sparse Grids}
In this section, we show how PIC may be used in concert with the combination technique to obtain a scheme with complexity that depends only logarithmically on dimension.  We proceed first by reinterpreting the usual PIC shape function $S$.  We then carry out a formal error and complexity analysis.  We conclude the section by summarizing the new algorithm.

\subsection{Reinterpreting the shape function}
Define $\bar{f} = f / \mathcal{N}$.  Note that $\bar{f}$ is non-negative and its phase space integral is unity.  It may thus be interpreted as a probability density.  Moreover, by definition, 
\begin{equation}
	\rho_e(\mathbf{x}) = q \int f \, d\mathbf{v} = q\mathcal{N} \int \bar{f}(\tilde{\mathbf{x}},\mathbf{v}) \delta(\mathbf{x} - \tilde{\mathbf{x}}) \, d\mathbf{v} d\tilde{\mathbf{x}} = Q \mathbb{E}_{\bar{f}(\tilde{\mathbf{x}},\mathbf{v})} [ \delta(\mathbf{x} - \tilde{\mathbf{x}}) ],
\end{equation}
where we have introduced notation for the expected value over the probability density $\bar{f}(\tilde{\mathbf{x}},\mathbf{v})$.  

Of course, the mean of a Dirac delta function is an entirely formal concept and impossible to compute from particle data.  With the introduction of a spatial grid, however, it is natural to introduce an approximate delta function - we choose
\begin{equation}
	\delta(\mathbf{x}) \approxeq \mathcal{S} (\mathbf{x}) \coloneqq \prod_{i=1}^{d_x} \frac{\tau \left(x_i / h_i \right)}{h_i}.
\end{equation}
We note that the error in approximating the delta function by $\mathcal{S}$ can be written as
\begin{equation} \label{Serrorexpansion}
	\rho_e(\mathbf{x}) - Q \mathbb{E}_{\bar{f}(\tilde{\mathbf{x}},\mathbf{v})}[\mathcal{S}(\mathbf{x} - \tilde{\mathbf{x}})] = C_1(h_x) h_x^2 + C_2(h_y) h_y^2 + C_3(h_x,h_y) h_x^2 h_y^2
\end{equation}
in two dimensions, while a directly analogous expression holds in 3-D.  This can be seen by simple Taylor expansion of $\bar{f}$ in position space.  As before, the $C_i$ are functions with a uniform upper bound and can be chosen to make this expression a true equality.  Already, this bears a promising resemblance to (\ref{uerr}).  Moreover, we note that any of the B-splines commonly used as shape functions for PIC will have an error of this form - indeed, any non-negative $\mathcal{S}$ can do no better than second order in this sense. The error given in \eqref{Serrorexpansion} may be interpreted as the grid error of the PIC method, as will be more apparent from our error analysis in Section \ref{sec:error}

Of course, in PIC, we cannot evaluate $\mathbb{E}_{\bar{f}(\tilde{\mathbf{x}},\mathbf{v})} [\mathcal{S}]$ exactly.  It is approximated by a sum over the particles, which in our probablistic interpretation are regarded as samples from the probablity density $\bar{f}$.  That is, we approximate $\rho_e$ at each grid point $\mathbf{x}_k$ by
\begin{equation} \label{reinterp_approx}
	\rho_{e} (\mathbf{x}_k) \approx Q\mathbb{E}_{\bar{f}(\tilde{\mathbf{x}},\mathbf{v})} [\mathcal{S}(\mathbf{x}_k - \tilde{\mathbf{x}})] \approx \frac{Q}{N_p} \sum_{p} \mathcal{S} (\mathbf{x}_k - \mathbf{x}_p) = \varrho_e(\mathbf{x}_k).
\end{equation}
For our $\mathcal{S}$, this approximation of $\rho_e$ is \textit{identical} to the standard viewpoint of (\ref{PICparadigm}). The difference between both sides of \eqref{reinterp_approx} may be interpreted as the statistical or particle error of the PIC method, as our analysis will clearly show in the next section.

\subsection{Formal error analysis}\label{sec:error}
Let us now assume that the initial particle states have been chosen by independent sampling from $\bar{f}(t=0)$ - a so-called \textit{noisy start}.  
%Strictly speaking, the particle states at later times are not independent random variables - they are coupled through the electric field.  However, when $N_p \gg 1$, this coupling is small, and particles may be regarded as approximately independent - this is essentially the molecular chaos assumption used in deriving kinetic equations.  
When $N_p \gg 1$, each particle has a small influence on the bulk field used to push the particles, meaning individual particle states remain approximately independent.  We may thus regard our approximation of $\rho_e(\mathbf{x}_k)$ as a random variable with mean $Q \mathbb{E}_{\bar{f}(\tilde{\mathbf{x}},\mathbf{v})} [\mathcal{S}(\mathbf{x}_k - \tilde{\mathbf{x}})]$ and variance given by $Q^2 \textrm{Var}_{\bar{f}(\tilde{\mathbf{x}},\mathbf{v})} [\mathcal{S}(\mathbf{x}_k - \tilde{\mathbf{x}})] / N_p$.

Taylor expanding $\bar{f}$ and evaluating simple integrals gives an expression for $\textrm{Var}_{\bar{f}(\tilde{\mathbf{x}},\mathbf{v})} [\mathcal{S}(\mathbf{x}_k - \tilde{\mathbf{x}})]$.  Namely, we have
\begin{equation}
	Q^2\textrm{Var}_{\bar{f}(\tilde{\mathbf{x}},\mathbf{v})} [\mathcal{S}(\mathbf{x}_k - \tilde{\mathbf{x}})] \approx \frac{4/9}{h_x h_y} Q\rho_e(\mathbf{x}_k)
\end{equation}
to leading order.  With this, we find that
\begin{equation}
	\rho_e(\mathbf{x}_k) - \varrho_e(\mathbf{x}_k) = C_1(h_x) h_x^2 + C_2(h_y) h_y^2 + C_3(h_x,h_y) h_x^2 h_y^2 + \xi_k,
\end{equation}
where $\xi_k$ is a random variable with 
\begin{equation} \label{varprops}
	\mathbb{E}[\xi_k] = 0, \qquad \mathrm{Var}[\xi_k] \approx \frac{4Q \rho_e(\mathbf{x}_k)}{9} \frac{1}{h_x h_y N_p}.
\end{equation}
As usual, a directly analogous expression holds in three dimensions.  

From here, we can directly see the grid and particle error scalings that were quoted in (\ref{stderrscalings}).  The introduction of $\mathcal{S}$ to approximate the Dirac delta function has introduced an $O(h^2)$ error, while approximation of an integral by a sum over $N_p$ particles has led to a random error of size $O(1/\sqrt{h_x h_y N_p})$.  Moreover, when $\varrho_e$ is extended to the entire computational domain by bilinear interpolation, exactly the same expression for the error holds because the interpolation error has the same form as the existing grid error.  

We are now in a position to leverage the sparse grid ideas outlined in section 2.1.  For notational simplicity, we once again assume the computational domain is $[0,1] \times [0,1]$, but all results are easily generalizable to an arbitrary rectangle.  Fix the particle number $N_p$ and let $\varrho_{i,j}$ denote the approximation of $\rho_e$ computed using $h_x = 2^{-i}$ and $h_y = 2^{-j}$ in (\ref{reinterp_approx}) and extended to the entire domain via bilinear interpolation.  Then 
\begin{equation}
	\varrho_n = \sum_{i+j = n+1} \varrho_{i,j} - \sum_{i+j=n} \varrho_{i,j}
\end{equation}
takes advantage of many of the same cancellations described in section 2.1.  In particular, 
\begin{equation}
\begin{split}
	\rho_e - \varrho_n &= h_n^2 \left\{C_1(h_n) + C_2(h_n) + \frac{1}{4} \sum_{i+j=n+1} C_3(h_x^i,h_y^j) - \sum_{i+j=n} C_3(h_x^i,h_y^j) \right\} \\ 
	&+ \sum_{i+j = n+1} \xi_{i,j} - \sum_{i+j=n} \xi_{i,j}
\end{split}
\end{equation}
As in standard versions of the combination technique, the term in braces is $O(n) = O(|\log h_n|)$.  Moreover, the mean-square size of the random error is
\begin{equation}
	\mathbb{E} \left[ \left( \sum_{i+j = n+1} \xi_{i,j} - \sum_{i+j=n} \xi_{i,j} \right)^2 \right] = O \left(n^2 (N_p h_n)^{-1} \right)
\end{equation}
since the left side contains $(2n-1)^2$ terms of the form
\begin{equation}
\begin{split}
	\mathbb{E} \left[ \xi_{i,j} \xi_{i',j'} \right] &\leq \left( \textrm{Var}[\xi_{i,j}] \textrm{Var}[\xi_{i',j'}] \right)^{1/2} \\
	&= O\left( (N_p h_x^i h_y^j)^{-1/2} \right) \cdot O \left( (N_p h_x^{i'} h_y^{j'})^{-1/2} \right) \\
	&= O \left( (N_p h_n)^{-1} \right)
\end{split}
\end{equation}
by the Schwarz inequality and the definitions of $h_x^i$ and $h_y^j$.  Thus, we find that the root-mean-square particle sampling error is now $O((N_p h_n)^{-1/2} |\log h_n |)$.  

Thus we find the overall error in the approximation of $\rho_e$ has the scaling 
\begin{equation}
	|\rho_e - \varrho_n| = \underbrace{O(h_n^2 |\log h_n|)}_{\textrm{Grid error}} + \underbrace{O\left( \frac{|\log h_n|}{\sqrt{h_n N_p}}\right)}_{\textrm{Particle error}}
\end{equation}
for 2-D sparse grids, and the only difference in 3-D is an extra power of $|\log h_n|$ in each term.  This gives rise to a sparse version of the standard PIC scalings in (\ref{stderrscalings}):
\begin{equation}
	\underbrace{\varepsilon \sim \Delta t^2}_{\substack{\text{Time-stepping}\\\text{error}}}, \qquad \underbrace{\varepsilon \sim h_n^2 |\log h_n|^{d_x - 1}}_{\text{Grid error}}, \qquad \underbrace{\varepsilon \sim |\log h_n|^{d_x-1} (N_p h_n)^{-1/2}}_{\text{Particle sampling error}}.
\end{equation}

If we again assume that particle steps dominate the complexity of the PIC scheme, then we find that the complexity scales as
\begin{equation}
	\kappa \sim \frac{N_p}{\Delta t} \sim \varepsilon^{-3} | \log \varepsilon|^{3(d_x - 1)}
\end{equation}
We see that, in contrast to standard PIC on regular grids, the complexity is \textit{nearly} independent of dimension, in that the complexity only increases by logarithmic powers of the desired error as dimension increases.  

The improvement over standard PIC can be intuitively understood in the following way.   The figure of merit for the statistical error in a PIC scheme is the number of particles \textit{per cell} - $N_p / N_c$, where $N_c$ denotes the number of cells.  Since $N_c$ scales inversely with the cell volume,  on a regular grid, $N_c \sim h^{-3}$ in 3-D.  However, on a 3-D sparse grid, $N_c \sim (4 h)^{-1}$.  We thus achieve \textit{many} more particles per cell, even with the total particle number fixed, by using a sparse grid.   

\subsection{Algorithm outline}
Using PIC in concert with the combination technique functions very much like standard PIC as outlined in section 2.2.  The analogous outline - in 2-D - is:
\begin{enumerate}[i.]
	\item Push particles exactly as in standard PIC.
	\item Assign to each particle a sequence of shape functions $\mathcal{S}_{i,j}(\mathbf{x} - \mathbf{x}_p) = \tau(2^i (x-x_p)) \tau(2^j (y-y_p)) / 2^{i+j}$, and approximate the overall electron charge density via
	\begin{equation} \label{sparsePICparadigm}
		\rho_e \approx \varrho_e \coloneqq \sum_{i+j = n+1} \varrho_{i,j} - \sum_{i+j=n} \varrho_{i,j}, 
	\end{equation}
	where $\rho_{i,j}$ is defined at grid points $\mathbf{x}_{k,\ell} = (k 2^{-i}, \ell 2^{-j})$ by 
	\begin{equation}
		\varrho_{i,j}(\mathbf{x}_{k,\ell}) = \frac{Q}{N_p} \sum_{p} \mathcal{S}_{i,j} (\mathbf{x}_p - \mathbf{x}_{k,\ell})
	\end{equation}
	and extended to the entire domain using bilinear interpolation.  
	\item Use some grid-based Poisson solver to compute $\varphi_{i,j}$ by solving $-\nabla^2 \varphi_{i,j} = \varrho_{i,j} + \rho_i$, then evaluate $\mathbf{E}_{i,j}$ by numerical differentiation of $\varphi_{i,j}$.  
	
	\item Evaluate $\mathbf{E}$ at the particle positions $\mathbf{x}_p$ via
	\begin{equation}
		\mathbf{E}(\mathbf{x}_p) = \sum_{i+j=n+1} \mathbf{E}_{i,j}(\mathbf{x}_p) - \sum_{i+j=n} \mathbf{E}_{i,j}(\mathbf{x}_p).
	\end{equation}  
	Use this to repeat step i.
\end{enumerate}

Step ii takes advantage of sparse grids to achieve an elevated number of particles per cell in our representation of the electron density.  Moreover, steps iii and iv also accelerate the ``field solve" in the standard manner of sparse grids as described in section 2.1.  

%%%%%%%%%%%%%%%%%%%%%%%%%%%%%%%%%%%%%%%%%%%%%%%%%%%%%%%%%%%%%%%%%%%%%%%%%%%%%%%%%%%%%%%%%%%%%%%%%%%%%%%%%%%%%%%%%%%%%%%%%%%%%%%%%%%%%%%%%%%%%%%%%%%%%%%%%%%%%%%%%%%%%%%%%%%%%%%%%

\section{Numerical Tests}
We work in dimensionless variables in which distance is measured in multiples of the Debye length $\lambda_D = \sqrt{\epsilon_0 T/q \rho_e}$, time in multiples of the inverse plasma frequency $\omega_p^{-1} = \sqrt{\epsilon_0 m_e/q \rho_e}$, and the electron mass $m_e$ and charge $q$ are each normalized to one.  In this initial study, we restrict our attention to a periodic box with side-lengths $L$.  

We perform three types of numerical test:
\begin{itemize}
	\item Linear Landau damping in 3-D
	\item Nonlinear Landau damping in 3-D
	\item Diocotron instability in 2-D
\end{itemize}
The purpose of the linear Landau damping test is to confirm the correct damping rate.  For the nonlinear Landau damping tests, we compare the accuracy and timing of the sparse PIC method to those of a standard, explicit PIC method.  The diocotron instability provides an example with a magnetic field and a chance to see sparse grids attempt to handle fine scale structure.

Throughout this section, we will refer to the number of simulated particles \textit{per cell}, which we denote by $P_c$.  For runs with standard PIC, the definition is standard:
\begin{equation}
	P_c = \frac{N_p}{N_c} = \frac{N_p h_x h_y}{L^2}
\end{equation}
in 2-D, and analogously in 3-D.  For the sparse-PIC runs, however, we count the total number of cells in \textit{all} the grids used in the combination technique.  Here, for an overall resolution of $2^n$ cells in each direction, one finds
\begin{equation}
	P_c = \frac{N_p}{n2^{n+1} + (n-1)2^n} = \frac{N_p 2^{-n}}{3n - 1}
\end{equation}
in 2-D, and similarly in 3-D.  For brevity, we also refer to a sparse grid whose maximum resolution in each coordinate is $2^n$ as being ``$2^n \times 2^n$" in 2-D and similarly in 3-D.  

We will measure accuracy in density $\rho_e$ and the electric field $\mathbf{E}$.  We do this by comparing to a highly resolved reference solution computed with standard PIC. The reported error in a given quantity $\psi$ will be given by
\begin{equation}
	\mathcal{E}(\psi) = \frac{\left\lVert \psi - \psi_{\textrm{ref}} \right\rVert_{L^2}}{\left\lVert \psi_{\textrm{ref}} \right\rVert_{L^2}} = \sqrt{\frac{\int \left| \psi - \psi_{\textrm{ref}} \right|^2 \, d\mathbf{x}}{\int \left| \psi \right|^2 \, d\mathbf{x}}},
\end{equation}
where $\psi_{\textrm{ref}}$ is the relevant reference solution, and integrals are taken over the computational domain. We compute these integrals with the trapezoidal rule. Since all the quantities are periodic and since we rely on an equispaced grid in all direction, quadratures with the trapezoidal rule give us high order accuracy \cite{trefethen2014exponentially}.  

When comparing the efficiency of sparse and regular PIC schemes, we measure both total computation time (in seconds) and total memory usage (in gigabytes).  The memory usage metric is of particular importance for large-scale applications to massively parallel architectures, where computations are increasingly memory bound.  In typical simulations, the dominant memory load is in storing the states (position and velocity) of the particles.  This is the usage we report, assuming double precision arithmetic.  

Finally, we note that energy conservation is an important consideration in PIC schemes of the type considered here.  We do not present detailed results, but do note that in all tests performed, the energy conservation properties were identical between regular- and sparse-PIC.  

\subsection{Notes on implementation and parallelization}
Before discussing detailed computational results, it is useful to discuss the implementation of the algorithm used.  We use the FFT to solve Poisson's equation for $\varphi$ and to differentiate it to find $\mathbf{E}$.  Additionally, to accommodate the enormous numbers of particles necessary, reference solutions are computed on the NYU HPC cluster `Mercer'.  However, all other computations are performed on a quad-core personal workstation except where specifically noted.  Regular PIC computations are performed entirely in serial, while sparse-PIC computations are `lightly parallelized' in the following sense.

In a shared memory context, the sparse-PIC algorithm is even more parallelizable than standard PIC.  To see this, consider the three operations that are performed on the particles in a PIC scheme:
\begin{itemize}
	\item Push particles
	\item Map grid data to particles
	\item Map particle data to grid
\end{itemize}
The first two operations are easily parallelizable since, if each thread is assigned to an individual particle, no two threads ever attempt to write to the same memory address.  However, as noted in \cite{buyukkecceci2013portable,stantchev2008fast} the mapping of particle data onto the grid is prone to memory collisions when threads are assigned particle-wise, as multiple particles may interact with a single grid cell.  This can be overcome through use of atomic operations, but a loss of performance results.  

In contrast, in the sparse-PIC method described here, particle data is interpolated onto multiple \textit{distinct} grids.  If each thread is assigned to a single grid used in the combination technique, then these computations can be performed simultaneously with no risk of memory collisions.  Our sparse-PIC implementation takes advantage of this added parallelism relative to standard PIC when computing the particle-to-grid map, but computations are otherwise performed in serial for fair comparison to standard PIC.

\subsection{Linear Landau damping}\label{sec:linearlandau}
In the limit of long wavelength, the damping rate $\gamma$ of a warm plasma wave is given in our dimensionless variables by \cite{krall1973principles}
\begin{equation}
	\gamma \approx \sqrt{\frac{\pi}{2}} \frac{\omega^2}{2k^3} \exp \left\{ -\frac{\omega^2}{2k^2} \right\}
\end{equation}
when the velocity distribution is Maxwellian.  Here, the frequency $\omega$ and wavenumber $k$ are related by the Bohm-Gross dispersion relation
\begin{equation}
	\omega^2 = 1 + 3 k^2.
\end{equation}
We introduce the shorthand
\begin{equation}
	g(x; \alpha, \beta) = 1 + \alpha \cos \frac{2 \beta \pi x}{L} 
\end{equation}
and choose the initial distribution
\begin{equation}
	f(t=0) = \frac{1}{(2\pi)^{3/2}} \exp \left\{ -\frac{v_x^2 + v_y^2 + v_z^2}{2} \right\} g\left(x; 0.05,1\right) g(y; 0.05, 1) g(z; 0.05,1),
\end{equation}
$L = 22$, $\Delta t = 1/20$ and final time $T = 25$ for these tests.   

\begin{figure}[h] 
\begin{center}
	\includegraphics[width=.6\textwidth]{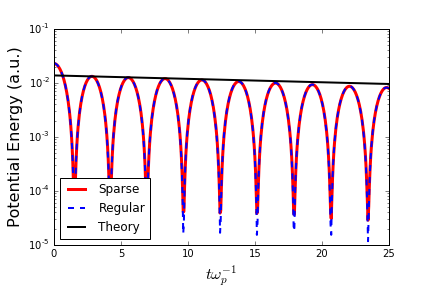}
	\caption{Potential energy as a function of time for the 3-D linear Landau damping test case described in section \ref{sec:linearlandau}.  Sparse (solid red) and regular (dashed blue) PIC simulations both agree well with predicted damping rate (black).}
	\label{PElandau}
\end{center}
\end{figure}

The results are shown in figure \ref{PElandau}.  Both the regular and sparse grid solutions use a $64 \times 64 \times 64$ grid resolution with $500$ particles per cell.  Both solutions agree well with each other and with the predicted damping rate.  The sparse solution is obtained with total particle number $N_p = 3.968 \times 10^6$, while the regular grid solution uses $N_p = 1.31 \times 10^8$ and requires approximately eight times the computation time of the sparse grid solution in our implementation.

\subsection{3-D Nonlinear Landau Damping}
For our nonlinear damping tests, we choose $L = 160$.  The initial distribution is given by
\begin{equation}
	f(t=0) = \frac{1}{(2\pi)^{3/2}} e^{-|v|^2/2} g\left(x; \, 0.2,4\right) g\left(y; \, 0.15, 3\right) g\left(z; \, 0.2, 4 \right).
\end{equation}

We use $\Delta t = 1/20$, and the reference solution is taken to be an ensemble average of 32 independent computations on a $128 \times 128 \times 128$ grid with $N_p = 1.5 \times 10^9$.  We perform standard- and sparse-PIC runs using overall grid resolutions of $16$, $32$, $64$, $128$, and with $P_c = 25, 50, 100, 400, 800$.  In addition, we run sparse tests at grid resolution $256$.  All sparse-PIC runs are performed on a personal workstation, but the memory requirements of regular-PIC necessitate running the most computationally demanding cases on a cluster.  

In all cases, we use $\Delta t = 1/20$ and we measure accuracy at the final time $T = 2.7$.  We begin by plotting the density at the final time for our reference solution, as well as sample sparse- and regular-PIC solutions in figure.  

\begin{figure}[h] 
\begin{center}
	\includegraphics[width=1.1\textwidth]{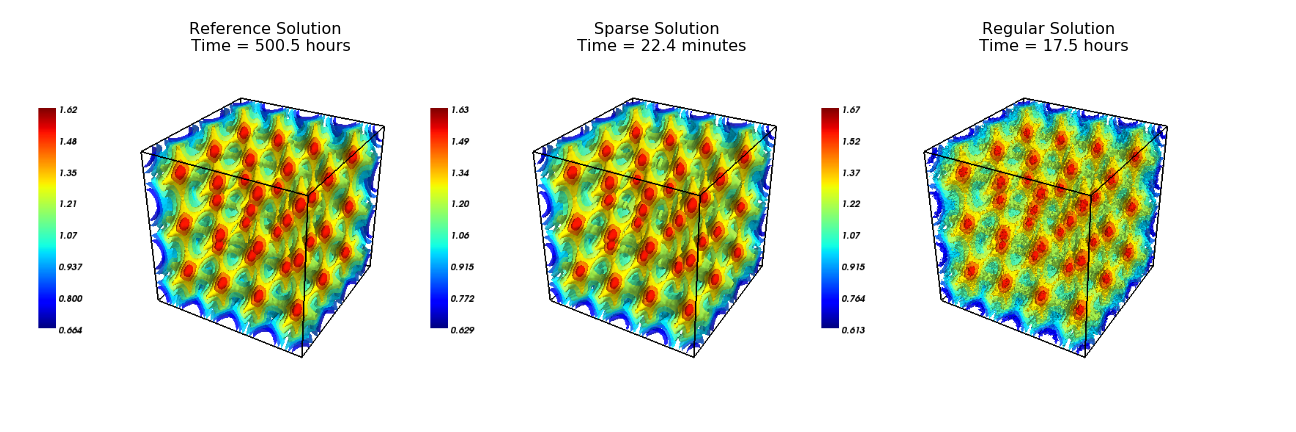}
	\caption{Comparison of density $\rho_e$ - normalized so that mean value is unity - between reference solution computed using regular PIC (left), sparse solution on $128 \times 128 \times 128$ grid with $P_c = 800$ (center), and regular-PIC solution on $128 \times 128 \times 128$ grid with $P_c = 800$ (right).  Approximate computation times are shown, revealing the dramatic speedup enabled by sparse grids.  Interestingly, even at the same $P_c$ value, sparse-PIC experiences visibly smaller statistical errors.}
	\label{3dcomparison}
\end{center}
\end{figure}

\begin{figure}[h] 
\begin{center}
	\includegraphics[width=.75\textwidth]{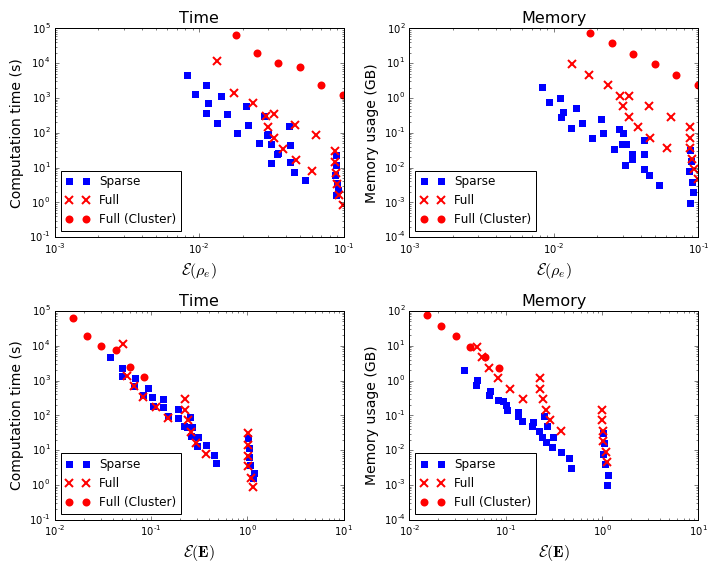}
	\caption{Computation time and memory usage for 3-D sparse (blue square) and full (red x) PIC runs.  Run times can be comparable, but sparse PIC consistently uses less memory, frequently by an order of magnitude or more.  This reduction in memory usage allows more accurate solutions for given computing resources. }
	\label{3deff}
\end{center}
\end{figure}

Next, we plot computation time and memory usage as functions of accuracy $\mathcal{E}(\rho_e)$ and $\mathcal{E}(\mathbf{E})$.  Results are shown in figure \ref{3deff}.  The advantages of the sparse approach are immediately evident in $\rho_e$, where for a given accuracy sparse-PIC is more than an order of magnitude faster and less memory intensive.  When measuring accuracy in the electric field, sparse-PIC is competitive in computation time while still generating reduction in memory usage.  

\begin{figure}[h] 
\begin{center}
	\includegraphics[width=.7\textwidth]{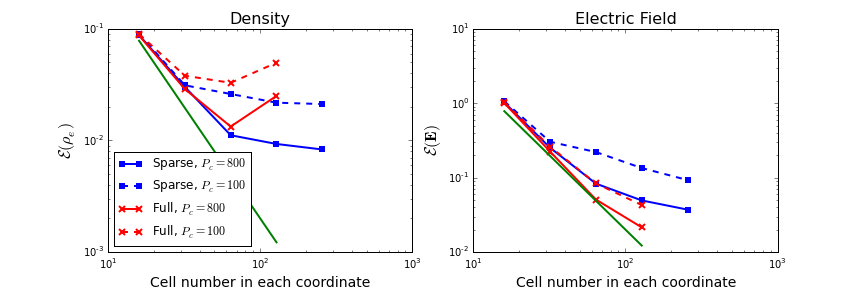}
	\caption{Accuracy in density $\rho_e$ and electric field $\mathbf{E}$ as functions of grid resolution.  }
	\label{3dconv}
\end{center}
\end{figure}

In addition, we plot $\mathcal{E}(\rho_e)$ and $\mathcal{E}(\mathbf{E})$ as functions of grid resolution in figure \ref{3dconv}.  As before, the sparse solution is superior when measuring $\rho_e$, but the electric field appears to be more sensitive to statistical error when sparse grids are used.  

\subsection{Diocotron Instability}
We consider here a variation on the diocotron instability, which is an instability that leads to the formation of vortices in electron plasmas with hollow density profiles confined by a uniform magnetic field \cite{aydemir1994unified,davidson2001physics,driscoll1990experiments}. The variation resides in the fact that in our case the electrons are immersed in a uniform, immobile and neutralizing background ion population, as described in \eqref{ESVlasov}, and we impose periodic boundary conditions. While this setup may seem somewhat artificial from a plasma physics point of view, it is often used to study vortex dynamics in neutral fluids \cite{melander1987axisymmetrization}, in which the diocotron instability is also observed \cite{dritschel1986nonlinear}, and may be viewed as an illustration of the Kelvin-Helmholtz instability \cite{driscoll1990experiments}. We expect fine scale structures to form as a result of this process, and are interested in the way our sparse grid scheme handles these structures.

To solve \eqref{ESVlasov} numerically, we choose magnetic field strength $B = 15$ oriented along the $z$-axis, $L = 22$ and
\begin{equation}
	f(t=0) = \frac{C}{2\pi} e^{-|v|^2/2} \exp \left\{ -\frac{(r-L/2)^2}{2 (0.03L)^2} \right\}, \quad r = \sqrt{(x - L/2)^2 + (y-L/2)^2}
\end{equation}
where $C$ is a normalization constant chosen to enforce $\int f dx dv = 1$.  As before, $\Delta t = 1/20$, while final time $T = 35$.  

\begin{figure}[h] 
\begin{center}
	\includegraphics[width=.8\textwidth]{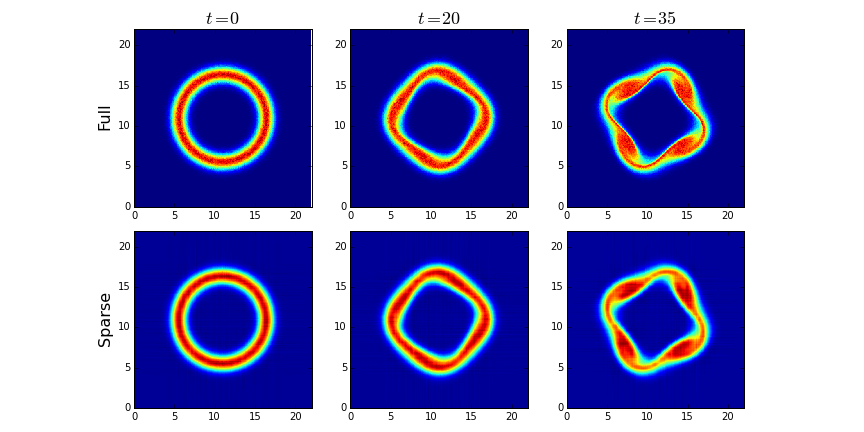}
	\caption{Two-dimensional test case of the diocotron instability at $t=0$ (left), $20$ (middle) and $35$ (right) using both regular PIC (top) and sparse-PIC (bottom).  Plots show density $\rho_e$.  Regular PIC used a $256 \times 256$ grid and $P_c = 40$, requiring $246$ seconds.  Sparse-PIC used a $1024 \times 1024$ grid and $P_c = 40$, requiring $225$ seconds.  }
	\label{2dcomp}
\end{center}
\end{figure}

We perform tests at $P_c = 20, 40, 80$.  For regular grids, we test grid resolutions of $64, 128, 256, 512$, while for sparse grids we additionally test $1024, 4096, 8192, 16384$.  Sample plots of $\rho_e$ are shown using both methods in figure \ref{2dcomp}.  

\begin{figure}[h] 
\begin{center}
	\includegraphics[width=.8\textwidth]{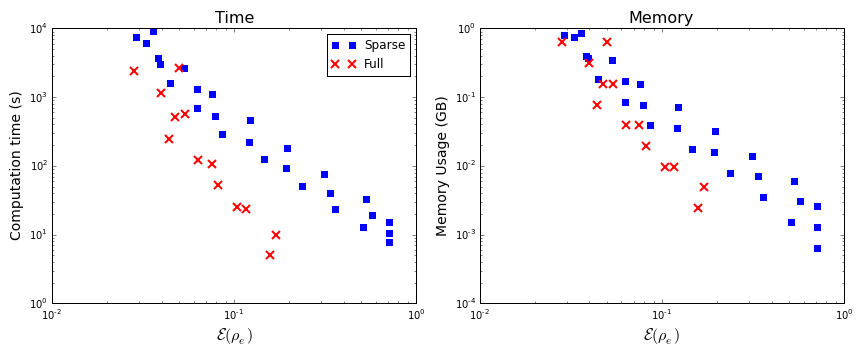}
	\caption{Computation time (left) and memory usage(right) as functions of accuracy for the 2-D diocotron example.}
	\label{2deff}
\end{center}
\end{figure}

We plot computation time and memory load as functions of $\mathcal{E}(\rho_e)$ in figure \ref{2deff}.  Evidently, regular-PIC is more efficient than sparse-PIC in 2-D for the parameters tested.  One can clearly see the improved asymptotic scaling of the computation time for sparse-PIC, but the so-called ``cross-over" point, at which sparse-PIC becomes more efficient than regular PIC occurs at larger scales than we are able to test here.  

It appears much of the error in the sparse-PIC runs arises from the tendency of sparse grids to ``smear" out the fine-scale structure developed by the solution - see the right-most portions of figure \ref{2dcomp}.  Higher-order sparse grid schemes and a better understanding of the interactions between sparse grids and the field solve remain promising options to rectify this shortcoming.  

\section{Conclusions}
In this initial study, we have presented the use of the sparse grid combination technique in concert with PIC methodology.  Although the method is still early in the development stage, initial results presented here demonstrate the method's potential to accelerate large scale kinetic plasma simulations. This is particularly true in three dimensions, for which the benefits of the sparse approach are immediately visible, even with our relatively simple implementation.  In the current state of the scheme, however, it is less clear whether sparse grids provide an advantage for 2-D simulations, except perhaps at very large grid sizes.

More work is needed to understand the interplay between statistical error and the field-solve part of the algorithm in the sparse grid context, as evidenced by figure \ref{3dconv}.  A possible alternative to the approach presented here is to compute the density $\rho_e$ on a sparse grid, but then to project it onto a regular grid for the purpose of the field-solve.  This is likely to reduce errors in the electric field due to the sparse grid representation. A drawback of this approach is that it would slow down the field-solve portion of the scheme to speeds commensurate with standard PIC, but in contexts where particle operations dominate the computation, this would be a small price to pay.   

There exist many other directions for future development.  Higher order interpolation schemes hold the potential to make sparse approaches even more attractive, even for two-dimensional simulations, by improving the ability to resolve fine structure.  Non-periodic boundaries, complex geometries, electromagnetic and collisional simulations are also subjects of ongoing work, with results to be reported at a later date.  

\section*{Acknowledgements}
The authors are particularly grateful to Bokai Yan, who is responsible for first introducing them to sparse grids.  Enlightening conversations were also had with Harold Weitzner and Russel Caflisch. The authors were supported by the U.S. Department of Energy, Office of Science, Fusion Energy Sciences under Award No. DE-FG02-86ER53223.  The second author emphasizes that the keys ideas for this article were contributed by the first author.

\vspace{4em}
\bibliographystyle{plain}
\bibliography{SherwoodPaper_arxiv.bbl}

\end{document}